\documentclass[12pt]{article}

\newlength{\abstwidth}
{\end{list}}
\newcounter{enumct}

\newcommand{\as}{\alpha_{\mathrm{s}}}
\newcommand{\aem}{\alpha_{\mathrm{em}}}
\renewcommand{\d}{\mathrm{d}}
\newcommand{\kT}{k_{\perp}}
\newcommand{\gast}{\gamma^*}

\newcommand{\pT}{p_{\perp}}

\begin{document}

\sloppy

\begin{flushright}
LU TP 00--52\\
hep-ph/0012185\\
December 2000
\end{flushright}

\vspace{2cm}

\begin{center}
{\LARGE\bf $\gamma\gamma$ Interactions}\\[3mm]
{\LARGE\bf from Real to Virtual Photons%
\footnote{to appear in the Proceedings of the Linear Collider %
Workshop 2000, Fermilab, October 24--28, 2000}}\\[10mm]
{\Large T. Sj\"ostrand\footnote{torbjorn@thep.lu.se}} \\[3mm]
{\it Department of Theoretical Physics,}\\[1mm]
{\it Lund University,}\\[1mm]
{\it S\"olvegatan 14A,}\\[1mm]
{\it S-223 62 Lund, Sweden}
\end{center}
 
\vspace{2cm}
 
\begin{center}
{\bf Abstract}\\[2ex]
\begin{minipage}{\abstwidth}
A `complete' framework for $\gamma\gamma / \gast\gamma / \gast\gast$
interactions is presented. The emphasis is on providing a model for
$\gamma\gamma$ physics at all photon virtualities, including
the difficult transition region $Q^2 \sim m_{\rho}^2$.
\end{minipage}
\end{center}

\vspace{1cm}

\noindent 
\rule{160mm}{0.5mm}

\vspace{1cm}

$\gamma\gamma$ physics is interesting in its own right, because of the 
challenges it offers to our understanding of complex QCD-related phenomena.
In addition, $\gamma\gamma$ events may offer a significant background to 
other kinds of physics studies, such as SUSY searches. This note 
summarizes the model recently presented in \cite{model}. It starts 
from the model for real photons in \cite{SaSmodel}, but further
develops this model and extends it also to encompass the physics of
virtual photons. The physics has been implemented in the \textsc{Pythia}
generator \cite{pythia}, so that complete events can be studied under 
realistic conditions.  

A first step in a calculation is the flux of incoming bremsstrahlung 
photons. Here a machinery has been set up for the flux of transverse 
and longitudinal photons as a function of the photon virtualities 
$Q_1^2$ and $Q_2^2$ and their longitudinal momentum fractions $y_1$ and 
$y_2$. Assuming isotropic azimuthal angles, $W^2$ can be calculated. 
The user can specify cuts on the range of these variables, in order to
restrict the generation to interesting events. 
The flux is convoluted with $\gast\gast$ cross sections dependent on 
$Q_1^2$, $Q_2^2$, $W^2$ and the polarization states. Beamstrahlung
involves only real photons, and is thereby considerably simpler. With
the flux of such photons given by some external program, already the
older machinery can cover this case.

Photon interactions are complicated since the photon wave function
contains so many components, each with its own interactions. To
first approximation, it may be subdivided into a direct and a resolved
part. (In higher orders, the two parts can mix, so one has 
to provide sensible physical separations between the two.)
In the former the photon acts as a pointlike particle, 
while in the latter it fluctuates into hadronic states.
These fluctuations are of $\mathcal{O}(\alpha_{\mathrm{em}})$, and so
correspond to a small fraction of the photon wave function, but this
is compensated by the bigger cross sections allowed in strong-interaction
processes. For real photons therefore the resolved processes dominate
the total cross section, while the pointlike ones take over for 
virtual photons. 
 
The fluctuations $\gamma \to q\overline{q} \, (\to \gamma)$ can be 
characterized 
by the transverse momentum $\kT$ of the quarks, or alternatively by some
mass scale $m \simeq 2 \kT$, with a spectrum of fluctuations 
$\propto {\d}\kT^2/\kT^2$. The low-$\kT$ part cannot be calculated 
perturbatively, but is instead parameterized by experimentally determined  
couplings to the lowest-lying vector mesons, $V = \rho^0$, $\omega^0$, 
$\phi^0$ and $J/\psi$, an ansatz called VMD for Vector Meson 
Dominance. Parton distributions are defined with a unit
momentum sum rule within a fluctuation \cite{SaSpdf}, giving rise
to total hadronic cross sections, jet activity, multiple interactions 
and beam remnants as in hadronic interactions. In interactions with a 
hadron or another resolved photon, jet production occurs by typical 
parton-scattering processes such as $qq' \to qq'$ or $gg \to gg$.

States at larger $\kT$
are called GVMD or Generalized VMD, and their contributions to the 
parton distribution of the photon are called anomalous. Given a dividing 
line $k_0 \simeq 0.5$~GeV to VMD states, the anomalous parton 
distributions are perturbatively calculable. The total cross section 
of a state is not, however, since this involves aspects of soft physics 
and eikonalization of jet rates. Therefore an ansatz is chosen where 
the total cross section of a state scales like $k_V^2/\kT^2$, where the 
adjustable parameter $k_V \approx m_{\rho}/2$ for light quarks. The 
spectrum of GVMD states is taken to extend over a range $k_0 < \kT < k_1$, 
where $k_1$ is identified with the $p_{\perp\mathrm{min}}(s)$ cut-off of 
the perturbative jet spectrum in hadronic interactions, 
$p_{\perp\mathrm{min}}(s) \approx 1.5$~GeV at typical 
energies \cite{pythia}. Above that range, the states are assumed to be 
sufficiently weakly interacting that no eikonalization procedure is 
required, so that cross sections can be calculated perturbatively 
without any recourse to Pomeron phenomenology. There is some 
arbitrariness in that choice, and some simplifications are required 
in order to obtain a manageable description.
 
A real direct photon in a $\gamma p$ collision can interact with the parton
content of the proton: $\gamma q \to qg$ (QCD Compton) and 
$\gamma g \to q\overline{q}$ (Boson Gluon Fusion). The $\pT$ in this 
collision is taken to exceed $k_1$, in order to avoid double-counting 
with the interactions of the GVMD states. In $\gamma\gamma$, the 
equivalent situation is called single-resolved 
(or direct$\times$resolved), where a direct photon interacts 
with the partonic component of the other, resolved photon. The 
$\gamma\gamma$ direct (or direct$\times$direct) process 
$\gamma\gamma \to q\overline{q}$ has 
no correspondence in $\gamma p$.

The space of $\gamma\gamma$ processes thus is three-dimensional, with axes
given by the $k_{\perp 1}$, $k_{\perp 2}$ and $p_{\perp}$ scales. Here 
each $k_{\perp i}$ is a measure of the virtuality of a fluctuation of 
a photon, and $p_{\perp}$ corresponds to the most virtual rung on 
the ladder between the two photons, possibly excepting the endpoint 
$k_{\perp i}$ ones. So, to first approximation, the coordinates along the 
$k_{\perp i}$ axes determine the characters of the interacting photons 
while $p_{\perp}$ determines the character of the interaction process. 
Double counting should be avoided by trying to impose a consistent 
classification. Thus, for instance, $p_{\perp} > k_{\perp i}$ 
with $k_{\perp 1} < k_0$ and $k_0 <k_{\perp 2} < k_1$ gives a hard 
interaction between a VMD and a GVMD photon, while 
$k_{\perp 1} > p_{\perp} > k_{\perp 2}$ with $k_{\perp 1} > k_1$ 
and $k_{\perp 2} < k_0$ is a single-resolved process
(direct$\times$VMD; with $p_{\perp}$ now in the parton distribution 
evolution).

If the photon is virtual, it has a reduced probability to fluctuate into 
a vector meson state, and this state has a reduced interaction probability.
This can be modelled by a traditional dipole factor
$(m_V^2/(m_V^2 + Q^2))^2$ for a photon of virtuality $Q^2$, where 
$m_V \to 2 \kT$ for a GVMD state. Putting it all together, the cross
section of the GVMD sector then scales like
\begin{equation}
\int_{k_0^2}^{k_1^2} \frac{{\d}\kT^2}{\kT^2} \, \frac{k_V^2}{\kT^2} \,
\left( \frac{4\kT^2}{4\kT^2 + Q^2} \right)^2 ~.
\end{equation}

For a virtual photon the DIS process $\gast q \to q$
is also possible, but by gauge invariance its cross section must
vanish in the limit $Q^2 \to 0$. At large $Q^2$, the single-resolved 
processes can be considered as the $\mathcal{O}(\as)$ correction to 
the lowest-order DIS process, but the single-resolved ones survive 
for $Q^2 \to 0$. There is no 
unique prescription for a proper combination at all $Q^2$, but we have
attempted an approach that gives the proper limits and minimizes 
double-counting. For large $Q^2$, the DIS $\gast \gamma$ cross section
is proportional to the structure function $F_2^{\gamma}(x, Q^2)$ with the
Bjorken $x = Q^2/(Q^2 + W^2)$. Since normal parton distribution 
parameterizations are frozen below some $Q_0$ scale and therefore do not
obey the gauge invariance condition, an ad hoc factor 
$(Q^2/(Q^2 + m_{\rho}^2))^2$ is introduced for the conversion from 
the parameterized $F_2^{\gamma}(x,Q^2)$ to a 
$\sigma_{\mathrm{DIS}}^{\gast \gamma}$:
\begin{equation}
\sigma_{\mathrm{DIS}}^{\gast \gamma} \simeq  
\left( \frac{Q^2}{Q^2 + m_{\rho}^2} \right)^2 \,
\frac{4\pi^2\aem}{Q^2} F_2^{\gamma}(x,Q^2) =
\frac{4\pi^2\aem Q^2}{(Q^2+m_\rho^2)^2} \,
\sum_{q,\overline{q}} e_{q}^2 \, x  q(x, Q^2) ~.
\label{sigDIS}
\end{equation}
Here $m_\rho$ is some non-perturbative hadronic mass parameter, for 
simplicity identified with the $\rho$ mass. 

In order to avoid double-counting between DIS and single-resolved events, 
a requirement $\pT > \max(k_1, Q)$ is imposed on the latter. In the 
remaining DIS ones, denoted lowest order (LO) DIS, thus $\pT < Q$. 
This would suggest a subdivision 
$\sigma_{\mathrm{LO\,DIS}}^{\gast \gamma} = 
\sigma_{\mathrm{DIS}}^{\gast \gamma} - 
\sigma_{\mathrm{1-res}}^{\gast \gamma}$, 
with $\sigma_{\mathrm{DIS}}^{\gast \gamma}$ given by 
eq.~(\ref{sigDIS}) and $\sigma_{\mathrm{1-res}}^{\gast \gamma}$ 
by the perturbative matrix elements. In the limit $Q^2 \to 0$, the 
DIS cross section is now constructed to vanish 
while the direct is not, so this would suggest 
$\sigma_{\mathrm{LO\,DIS}}^{\gast \gamma} < 0$. However, here we 
expect the correct answer not to be a negative number but an 
exponentially suppressed one, 
by a Sudakov form factor. This modifies the cross section: 
\begin{equation}
\sigma_{\mathrm{LO\,DIS}}^{\gast \gamma} = 
\sigma_{\mathrm{DIS}}^{\gast \gamma} -
\sigma_{\mathrm{1-res}}^{\gast \gamma}
~~ \longrightarrow ~~ 
\sigma_{\mathrm{DIS}}^{\gast \gamma} \; \exp \left( - \frac{%
\sigma_{\mathrm{1-res}}^{\gast \gamma}}%
{\sigma_{\mathrm{DIS}}^{\gast \gamma}} \right) \;. \label{eq:LODIS}
\end{equation}
Since we here are in a region where the DIS cross section is no longer the 
dominant one, this change of the total DIS cross section is not essential.

The space of $\gast\gast$ processes is now five-dimensional: $Q_1$, $Q_2$,
$k_{\perp 1}$, $k_{\perp 2}$ and $p_{\perp}$. As before, an effort is 
made to avoid double-counting, by having a unique classification of each 
region in the five-dimensional space. Some double-counting remain in the 
region of large $x \approx Q^2/(Q^2 + W^2)$, where an ad hoc suppression 
factor $(1-x)^3$ is introduced for the resolved components.  
 
In total, our ansatz for $\gast\gast$ interactions at all $Q^2$ contains 
13 components: 9 when two VMD, GVMD or direct photons interact, as is 
already allowed for real photons, plus a further 4 where a `DIS photon' 
from either side interacts with a VMD or GVMD one. With the label 
resolved used to denote VMD and GVMD, one can write
\begin{eqnarray}
\sigma_{\mathrm{tot}}^{\gast\gast} (W^2, Q_1^2, Q_2^2) & = &
\sigma_{\mathrm{DIS}\times\mathrm{res}}^{\gast\gast} \; 
\exp \left( - \frac{\sigma_{\mathrm{dir}\times\mathrm{res}}^{\gast\gast}}%
{\sigma_{\mathrm{DIS}\times\mathrm{res}}^{\gast\gast}} \right) +
\sigma_{\mathrm{dir}\times\mathrm{res}}^{\gast\gast} 
\nonumber \\
 & + & \sigma_{\mathrm{res}\times\mathrm{DIS}}^{\gast\gast} \; 
\exp \left( - \frac{\sigma_{\mathrm{res}\times\mathrm{dir}}^{\gast\gast}}%
{\sigma_{\mathrm{res}\times\mathrm{DIS}}^{\gast\gast}} \right) +
\sigma_{\mathrm{res}\times\mathrm{dir}}^{\gast\gast}   \\
 & + & \sigma_{\mathrm{dir}\times\mathrm{dir}}^{\gast\gast}
+ \left( \frac{W^2}{Q_1^2 + Q_2^2 + W^2} \right)^3 \;
\sigma_{\mathrm{res}\times\mathrm{res}}^{\gast\gast} \nonumber
\end{eqnarray}
Most of the 13 components in their turn have a complicated internal 
structure, as we have seen. 

An important note is that the $Q^2$ dependence of the DIS and direct 
photon interactions is implemented in the matrix element expressions, 
i.e. in processes such as $\gast\gast \to q\overline{q}$ or 
$\gast q \to q g$ the photon virtuality explicitly enters. This is 
different from VMD/GVMD, where dipole factors are used to reduce the 
total cross sections and the assumed flux of 
partons inside a virtual photon relative to those of a real one, but 
the matrix elements themselves contain no dependence on the virtuality
either of the partons or of the photon itself. 
Typically results are 
obtained with the SaS~1D PDF's for the virtual transverse photons
\cite{SaSpdf}, since these are well matched to our framework, e.g.
allowing a separation of the VMD and GVMD/anomalous components. 
Parton distributions of virtual longitudinal photons are by default
given by some $Q^2$-dependent factor times the transverse ones.
The new set by Ch\'yla \cite{Chyla} allows more precise modelling 
here, but first indications are that many studies will not be sensitive 
to the detailed shape.

Some first studies with this new framework look promising \cite{Sharka},
but much further work remains to assess its usefulness.

\end{document}